\documentclass[prl,twocolumn,preprintnumbers,amsmath,amssymb]{revtex4-1}  

\usepackage{graphicx}
\usepackage{dcolumn}
\usepackage{bm}
\usepackage{multirow}
\usepackage{color}

\usepackage[colorlinks=true]{hyperref}

\begin{document}
\title{ Exciton Valley Dynamics probed by Kerr Rotation in WSe$_2$ Monolayers}

\author{C.R. Zhu$^1$}
\author{K. Zhang$^1$}
\author{M. Glazov$^2$}
\author{B. Urbaszek$^3$}
\author{T. Amand$^3$}
\author{Z. W. Ji$^4$}
\author{B.L. Liu$^1$}
\email{blliu@aphy.iphy.ac.cn}
\author{X. Marie$^3$}
\email{marie@insa-toulouse.fr}
\affiliation{%
$^1$Beijing National Laboratory for Condensed Matter Physics, Institute of Physics, Chinese Academy of Sciences, PO Box 603, Beijing 100190, P.R. of China\\
$^2$Ioffe Physical-Technical Institute of the RAS, 194021 St. Petersburg, Russia\\
$^3$Universit\'e de Toulouse, INSA-CNRS-UPS, LPCNO, 135 Av. de Rangueil, 31077 Toulouse, France\\
$^4$School of Physics, Shandong University, Jinan 250100, China}


\begin{abstract}
We have experimentally studied the pump-probe Kerr rotation dynamics in WSe$_2$ monolayers. This yields a direct measurement of the exciton valley depolarization time $\tau_v$. At $T= 4$~K, we find  $\tau_v\approx 6$~ps, a fast relaxation time resulting from the strong electron-hole Coulomb exchange interaction in bright excitons. The exciton valley depolarization time decreases significantly when the lattice temperature increases with $\tau_v$  being as short as 1.5~ps at 125~K. The temperature dependence is well explained by the developed theory taking into account exchange interaction and fast exciton scattering time on short-range potential.
\end{abstract}


\maketitle

Monolayers of transition metal dichalcogenides (TMDC), such as MoS$_2$ and WSe$_2$, are two-dimensional (2D) semiconductors with strong light absorption and emission associated to direct optical transitions \cite{Butler:2013a,Mak:2010a,Splendiani:2010a}. The optical properties of these 2D crystals are strongly influenced by excitons, Coulomb bound electron-hole pairs, with experimentally determined binding energies of up to  0.6 eV ($\sim 1/3$ of the optical bandgap $\sim 1.7$~eV)~\cite{Cheiwchanchamnangij:2012a,Ye:2014a,Zhu:2014a,Klots:2014a,Ugeda:2014a,He:2014a,Wang:2014a}. Due to the combined effect of inversion symmetry breaking and strong spin-orbit interaction, the interband transitions are governed by chiral selection rules which allow efficient optical initialization of an electron-hole pair in a specific $K$-valley in momentum space \cite{Xiao:2012a, Cao:2012a,Mak:2012a,Sallen:2012a,Zeng:2012a}. The circular polarization ($\sigma^+$ or $\sigma^-$) of the absorbed or emitted photon can be directly associated with selective carrier excitation in one of the two non-equivalent valleys, $K_+$ or $K_-$, respectively, formed at the edges of the Brillouin zone. Hence, the spin state of an exciton  $S_z=\pm1$ is correlated with its valley state $K_\pm$.
Recent time-resolved studies demonstrate photoluminescence (PL) lifetimes in the picosecond range indicating high exciton oscillator strengths \cite{Korn:2011a,Lagarde:2014a, Wang:2014b}. Pump-probe absorption and reflectivity measurements in monolayer (ML) MoS$_2$ have also shown polarization decay times in the picosecond range \cite{Shi:2013b,Wang:2013d,Mai:2014a} corresponding to fast relaxation of the valley polarization, which is surprising as the valley degree of freedom is expected to be protected by the considerable single particle spin splittings in the valence and conduction bands \cite{Liu:2013a}.

In this work we use for the first time a powerful technique, namely time-resolved Kerr Rotation (TRKR)~\cite{Dyakonov:2008a}, to investigate exciton dynamics in ML WSe$_2$. TRKR allows, in contrast to time-resolved PL spectroscopy, to address the spin states of both photocreated and resident carriers polarized by a $\sigma^+$ or $\sigma^-$ polarized pump laser. We detect spin-Kerr signal for the neutral exciton, well separated from the charged exciton signal, and uncover a valley relaxation time of $\tau_v=6$~ps at $T=4$ K,  which we attribute to the strong Coulomb exchange interaction between the electron and the hole. The strongly bound excitons in ML WSe$_2$~\cite{Wang:2014b} give access in TRKR to a novel temperature-dependent regime of exciton spin dynamics: in standard quasi-2D semiconductor systems like GaAs quantum wells excitons are ionized as the temperature increases since their binding energy is small ($\sim10$~meV)~\cite{Vinattieri:1994a,Dareys:1993a,Maialle:1993a}. The temperature dependence of the exciton spin dynamics is therefore inaccessible. This is in contrast to very robust excitons in ML WSe$_2$, where we can investigate this evolution. We measure a temperature induced decrease of $\tau_v$ that we interpret in terms of the temperature dependence of the exchange interaction induced exciton spin relaxation. Our theory describes very satisfactorily the experimental data.  Moreover, we found no evidence of transfer of spin/valley polarization to resident carriers in WSe$_2$ ML contrary to similar time-resolved Kerr experiments performed in III-V or II-VI semiconductors~\cite{Awschalom:1997a,Glazov:2012a}.

\begin{figure}[t]
\includegraphics[width=0.4\textwidth]{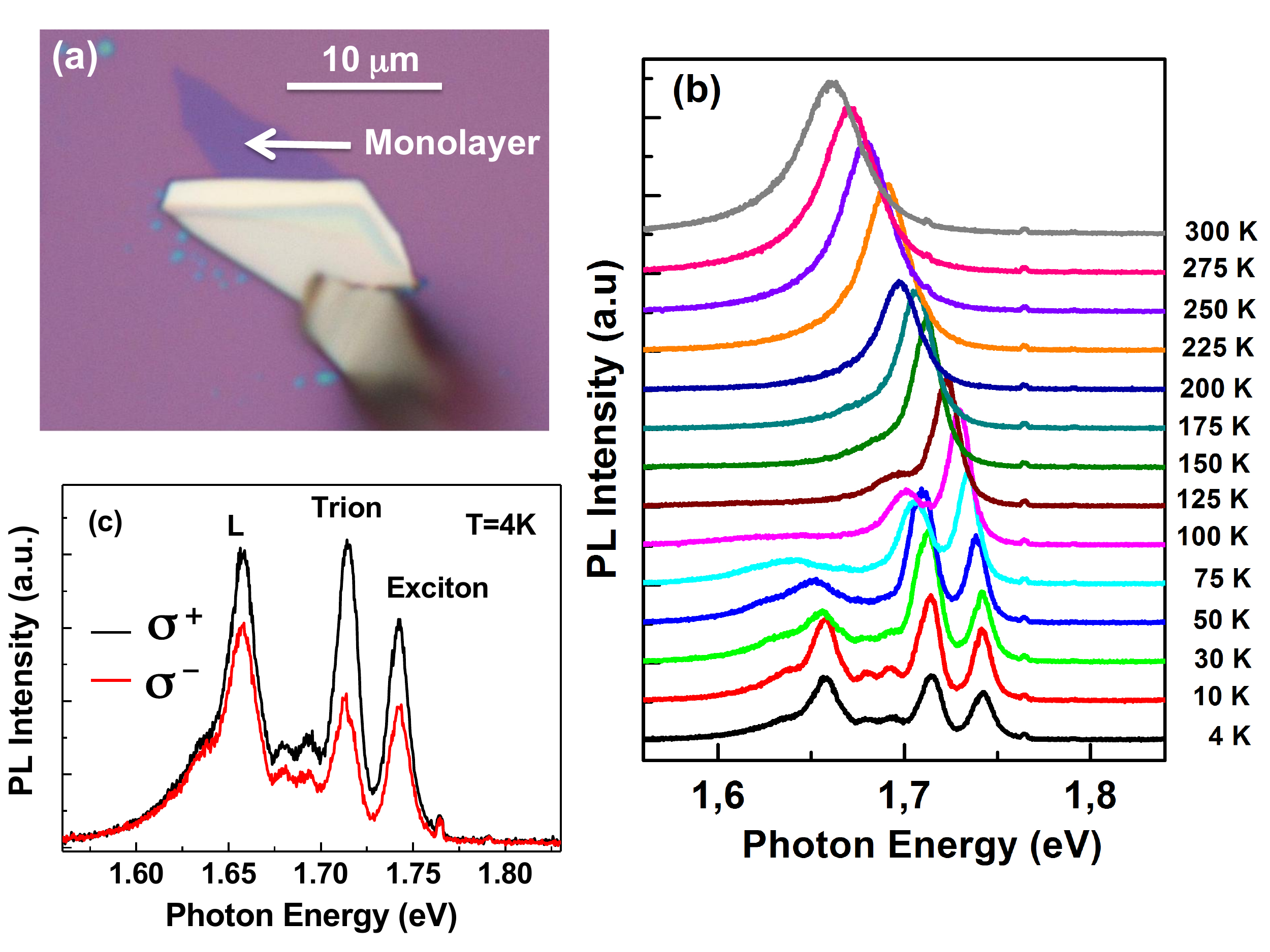}
\caption{\label{fig:fig1}   (a)	Optical reflection image of the sample indicating the investigated WSe$_2$ ML flake (the white zone corresponds to multiple layers). (b)	Temperature dependence of the WSe$_2$ ML PL spectra. 
(c)	Right ($\sigma^+$) and left ($\sigma^-$)  circularly polarized PL components at $T=4$~K under \emph{cw} $\sigma^+$ polarized  He-Ne laser excitation ($E_l=1.96$ eV). }
\end{figure} 

\begin{figure}
\includegraphics[width=0.4\textwidth]{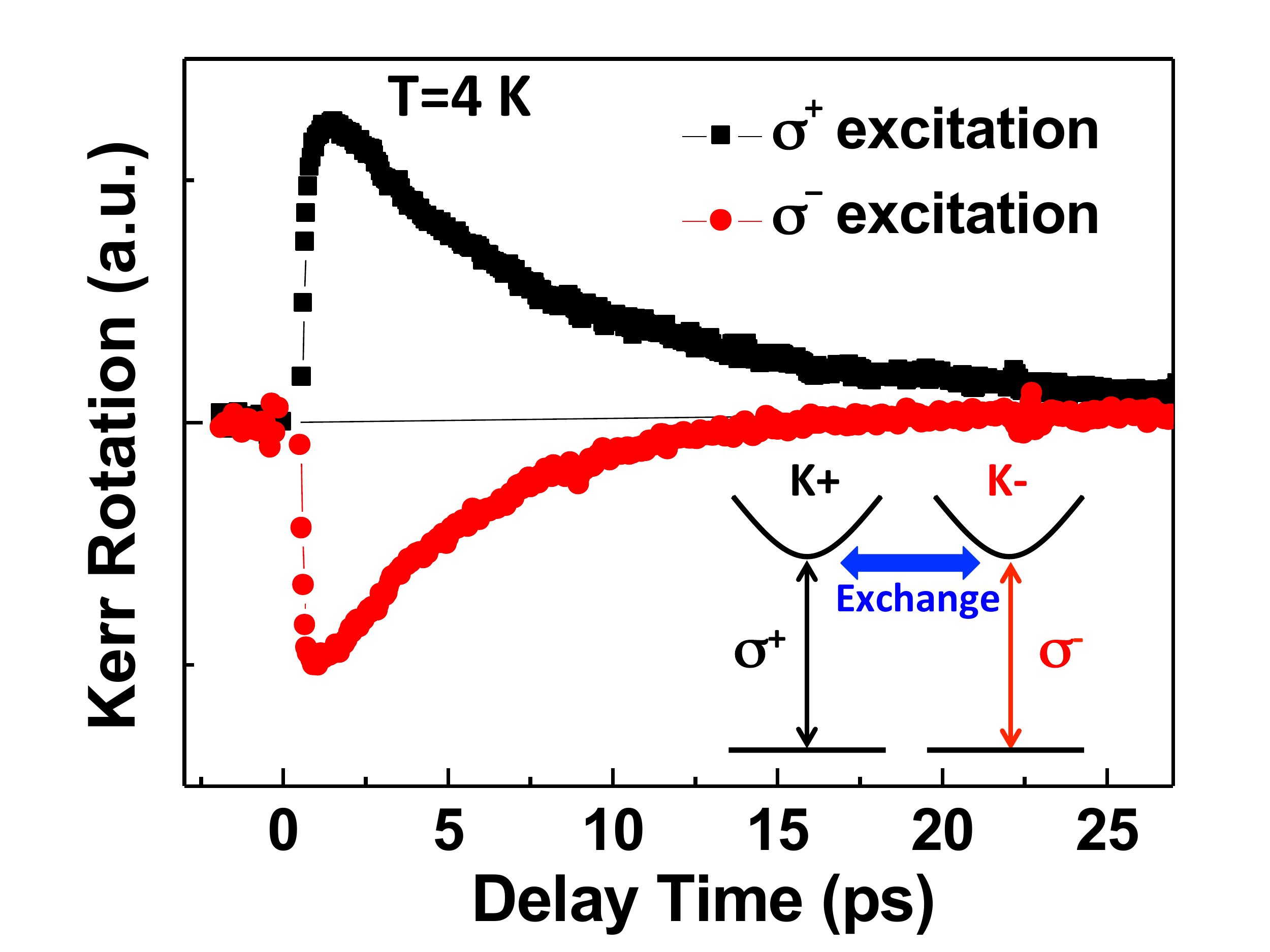}
\caption{\label{fig:fig2} Kerr rotation dynamics at $T=4$~K for $\sigma^+$ and $\sigma^-$ pump beam. The laser excitation energy is $E_l=1.735$~eV. Inset: schematics of the optical selection rules of the  excitons in $K_\pm$ valleys and their coupling induced by the long-range exchange interaction.}
\end{figure} 

The investigated monolayer WSe$_2$ flakes are obtained by micro-mechanical cleavage of a bulk WSe$_2$ crystal (from SPI Supplies, USA) on 90 nm SiO$_2$ on a Si substrate. The 1ML region is identified by optical contrast, Fig.~\ref{fig:fig1}a, and very clearly in PL spectroscopy \cite{Mak:2010a}. 
The sample is excited near normal incidence with degenerate pump and delayed probe pulses from a  mode-locked Ti:Sapphire laser ($\sim$120 fs pulse duration, 76 MHz repetition frequency). The laser beams are focused to a spot size of $\sim$ 5$\mu$m, and the pump and probe beams have average power of 300 $\mu$W and 30 $\mu$W, respectively. This corresponds to a typical pump-generated exciton density of about 10$^{12}$ cm$^{-2}$. The circularly polarized pump pulse incident normal to the sample creates spin-polarized electronic excitations with the spin vector perpendicular to the flake plane. The temporal evolution of the carrier spins is recorded by measuring the Kerr rotation angle $\theta(\Delta t)$ of the reflected linearly polarized probe pulse while sweeping the delay time $\Delta t$, which correspond to the net spin component normal to the sample plane~\cite{Liu:2007a,Dyakonov:2008a,Glazov:2012a}. Continuous-wave (\emph{cw}) micro-PL is performed with a laser excitation energy $E_l=1.96$~eV with a standard monochromator coupled to a CCD detection system.
Figure~\ref{fig:fig1}b displays the temperature dependence of the PL spectra from $T=4$~K to 300 K. Similarly to previous studies, three features can be identified at $T=4$~K \cite{Jones:2013a,Wang:2014b}: the emission peaks at $E=1.742$~eV and $E=1.714$~eV correspond to the recombination of neutral exciton and charged exciton (trion) respectively, Fig.~\ref{fig:fig1}c. Considering the commonly observed residual \emph{n}-type doping \cite{Radisavljevic:2013a,Wang:2014b}, the trion charge is assumed to be negative but this assumption is not critical for the present study and the discussion below.
The identification of these transitions is based on the emission polarization analysis \cite{Jones:2013a,Wang:2014b}. 
Under $\sigma^+$ polarized excitation light, the exciton and trion PL peaks are characterized  a significant circular polarization degree  (typically 33\% and 23\% respectively at $T=4$~K), note the different intensities of the $\sigma^+$ and $\sigma^-$ PL components in Fig.~\ref{fig:fig1}c,
 which demonstrate the optical initialization of valley polarization \cite{Xiao:2012a, Cao:2012a,Mak:2012a,Sallen:2012a,Zeng:2012a}. In contrast, only the exciton line exhibits linear polarization degree following a linearly polarized light excitation (not shown) as a consequence of the creation of a coherent superposition of valley states \cite{Jones:2013a}. The clear separation by ~30 meV of the trion and the neutral exciton  in WSe$_2$ ML is a major advantage compared to 
state-of-the-art MoS$_2$ ML samples where the two lines can not be resolved \cite{Plechinger:2014a}. Below the trion emission several emission peaks are observed at low temperature (labelled L in Fig.~\ref{fig:fig1}c); these peaks already observed in MoS$_2$ or WSe$_2$ MLs \cite{Xiao:2012a,Cao:2012a,Mak:2012a,Wang:2014b} have been assigned to localized exciton complexes. In this work we focus on the temperature dependence of the neutral exciton dynamics. As shown in Fig.~\ref{fig:fig1}b, the exciton emission remains strong as the temperature increases
(we observe clearly the red shift of the line) whereas both the trion and localized state emission vanish for $T\gtrsim 100$~K.

\begin{figure}
\includegraphics[width=0.5\textwidth]{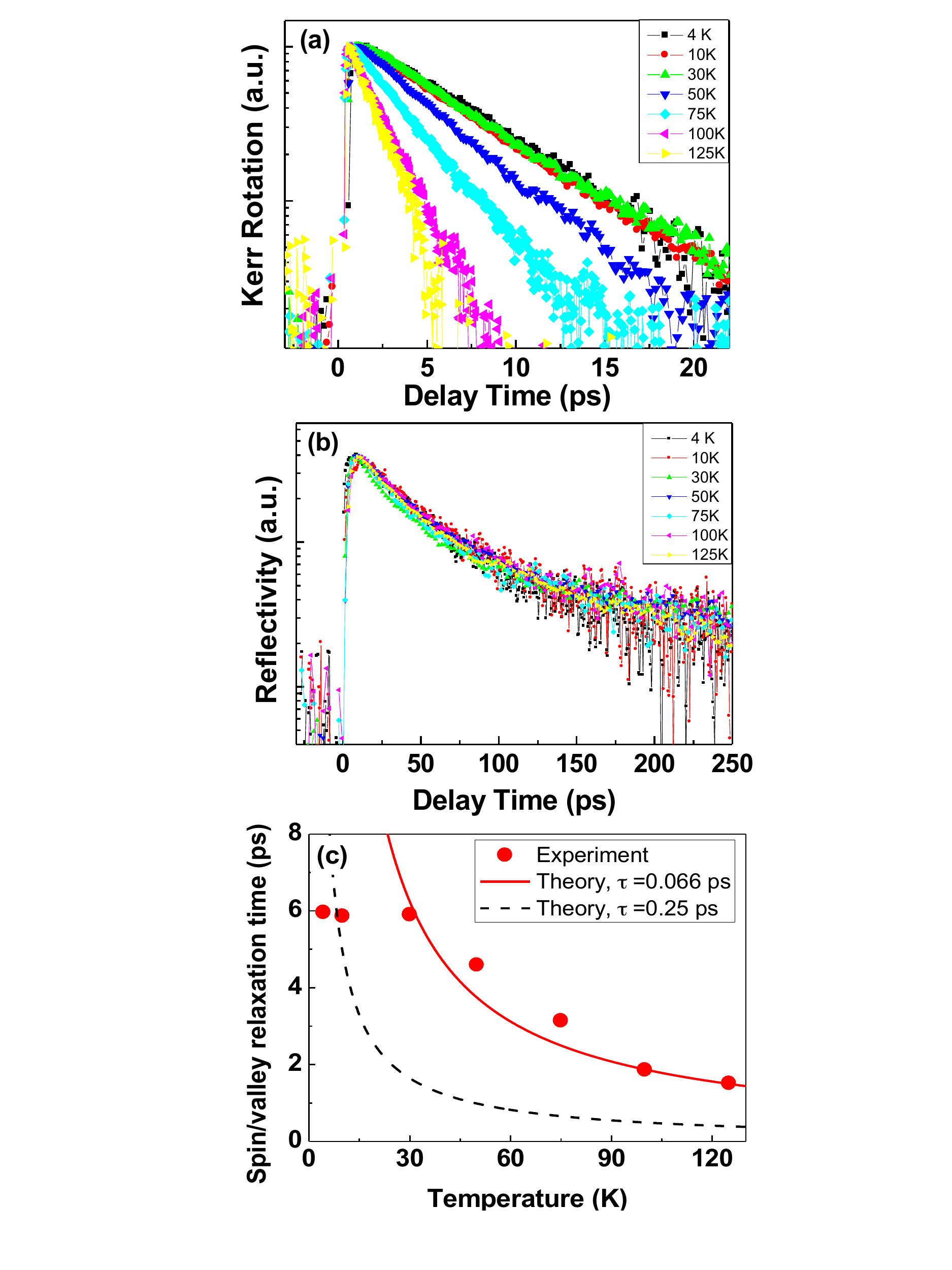}
\caption{\label{fig:fig3} (a) Kerr rotation dynamics after a $\sigma^+$ polarized pump 
pulse for different lattice temperatures;
(b)	Transient reflectivity dynamics for different temperatures. The laser excitation energies are identical to the ones used in (a), see text;
(c)	Temperature dependence of the measured (symbol) and calculated after Eq.~\eqref{tau:zz} (solid and dashed lines) exciton valley polarization relaxation time, see text for details.
 }
\end{figure}

Figure~\ref{fig:fig2}  shows the Kerr rotation dynamics $\theta(\Delta t)$ measured at 4 K for both $\sigma^+$ and $\sigma^-$ polarized pump pulses. The pump energy $E_{l}=  1.735$~eV is set to the maximum of the Kerr signal, which is very close to the neutral exciton transition identified in the PL spectra Fig.~\ref{fig:fig1}c. The observed sign reversal of the Kerr signal in Fig.~\ref{fig:fig2} at the reversal of pump helicity is a consequence of the optical initialization of the $K_+$ and $K_-$ valley, respectively. We have measured in the same conditions the transient reflectivity using linearly cross-polarized pump and probe pulses. As shown in Fig.~\ref{fig:fig3}b, the reflectivity decay time is about ten times longer than the one observed in TRKR. Thus the mono-exponential decay time $\tau_v=(6 \pm 0.1)$~ps of the Kerr rotation dynamics at $T=4$~K in Fig.~\ref{fig:fig2}, probes directly the fast exciton valley depolarization. In agreement with recent numerical estimations, it results from the strong long-range exchange interaction \cite{Glazov:2014a,Yu:2014a}. Due to limited time resolution, previous investigations by time-resolved PL spectroscopy could not yield the measurement of the exciton valley depolarization time \cite{Lagarde:2014a,Wang:2014b}. Kerr rotation dynamics used here is characterized by a higher time resolution ($\sim$100 fs) and allows us a strictly resonant excitation of the exciton. 

Figure~\ref{fig:fig3}a displays the variation of the TRKR dynamics as a function of the temperature. For each temperature, the laser excitation energy is set at the maximum Kerr rotation signal which follows well the energy of the neutral exciton PL shown in Fig.~\ref{fig:fig1}b.
For $T\gtrsim 30$~K, we observe a clear decrease of the exciton valley polarization decay time $\tau_v$ down to 1.5 ps at $T=125$~K; for higher temperatures the signal-to-noise ratio is too small to get reliable data. We stress that the transient reflectivity measurements performed in the same temperature range exhibit much longer decay times with very weak temperature dependence, Fig.~\ref{fig:fig3}b~\cite{Shi:2013b}. We have also investigated the excitation power dependences of the exciton dynamics. In the studied power range corresponding to variation of exciton density from $\sim 1.5 \times 10^{11}$ to $10^{12}$~cm$^{-2}$, both the TRKR and reflectivity dynamics do not depend on the exciton photo-generated density within the experimental accuracy (not shown). This demonstrates that the exciton-exciton interactions play a minor role in exciton valley dynamics  
presented in Figs. \ref{fig:fig2} and \ref{fig:fig3}.

In order to describe quantitatively the experimental findings we note that the Kerr rotation angle $\theta$ is proportional to the total spin of the exciton ensemble $\bm S=\sum_{\bm K} \bm S_{\bm K}$ (the mechanisms of Kerr rotation by exciton spins in WSe$_2$ MLs are similar to that in quasi-2D semiconductors and can be described as in Ref.~\cite{Glazov:2012a}); $\bm S_{\bm K}$ is the spin distribution function (with $S_z=\pm 1$ corresponding to $K_{\pm}$ valleys) and $\bm K$ is the wavevector of exciton. The spin/valley dynamics of excitons is governed by the long-range exchange interaction between an electron and a hole~\cite{Glazov:2014a,Yu:2014a} which acts as an effective magnetic field $\bm \Omega_{\bm K} = \alpha K (\cos{2\vartheta}, \sin{2\vartheta})$ on the exciton spin. Here $\vartheta$ is the polar angle of $\bm K$ and the constant $\alpha$ depends on the oscillator strength of exciton transition and system geometry. Following Ref.~\cite{Glazov:2014a} for the system ``vacuum --  1ML of WSe$_2$ -- substrate'' we obtain, assuming the same background refractive index $n$ of the ML and the substrate, $\alpha =  c\Gamma_0  (n+1)/[(n^2+1)\omega_0]$, where $\omega_0$ is the exciton resonance frequency, $\Gamma_0$ is its radiative lifetime~\footnote{In the electrodynamical approach of Ref.~\cite{Glazov:2014a} to the exciton LT-splitting, the reflection at the boundary ``vacuum -- 1 ML'' yields the replacement $\Gamma_{0,\alpha} \to \Gamma_{0,\alpha}(1+r_\alpha)$, where $\alpha=s,p$ and $r_\alpha$ are given by the Fresnel formula with background refractive index $n$.}. This effective field causes spin precession of excitons which is randomized by the scattering and described by the kinetic equation~\cite{Maialle:1993a,Glazov:2014a}
${\partial \bm S_{\bm K}}/{\partial t} + S_{\bm K} \times \bm \Omega_{\bm K}  = \bm Q\{ \bm S_{\bm K}\}$,  
where $Q\{ \bm S_{\bm K}\}$ is the collision integral.  Assuming that the excitons are thermalized and the scattering is caused by short-range potential we obtain for spin/valley relaxation rate $\tau_{zz}^{-1}$
\begin{equation}
\label{tau:zz}
\frac{1}{\tau_{zz}} = \frac{2\alpha^2 \tau M k_B T}{\hbar^2},
\end{equation}
where $M=0.67m_0$ is the exciton mass~\cite{Wang:2014b},
$\tau$ is the scattering time. Equation~\eqref{tau:zz} is valid provided that the radiative lifetime of excitons exceeds $\tau_{zz}$, $k_B T \tau/\hbar \gg 1$ and $\alpha^2 M k_B T \tau^2/\hbar^2 \ll 1$. The second condition means that the average kinetic energy must be larger than the homogeneous broadening and the latter that the spin relaxation occurs in the spin diffusive regime. Figure~\ref{fig:fig3}c shows the experimentally measured exciton polarization decay times (points are extracted from the monoexponential decay presented in Fig.~\ref{fig:fig3}a) and theoretical calculations carried out for $\Gamma_0 = 0.16$~ps$^{-1}$ (corresponding to the radiative lifetime of the states in the light cone $1/(2\Gamma_0) = 3$~ps, a value consistent with measurements \cite{Wang:2014b}) and $n=\sqrt{10}$. Qualitatively, the drop of the exciton spin/valley relaxation time when $T$ increases can be well explained by the increase with the temperature of the effective magnetic field $\bm \Omega_{\bm K}$, which makes spin precession and decoherence faster, Eq.~\eqref{tau:zz}. The solid and dashed lines in Fig.~\ref{fig:fig3}c correspond to the calculated exciton spin/valley relaxation time for  two different scattering times $\tau=0.066$~ps and $\tau=0.25$~ps, respectively. The latter value of $\tau$ corresponds to exciton energy uncertainty equivalent to $30$~K corresponding to Ioffe-Regel criterion of delocalization where the product of the thermal wavevector $k_T=(2Mk_BT)^{1/2}/\hbar$ and the mean free path $l=\tau \hbar k_T/M$ is on the order of $1$. Remarkably, we observe a  nice agreement between the calculated and measured exciton relaxation times in Fig.~\ref{fig:fig3}c for $T> 30$~K  with the scattering time $\tau=0.066$~ps, but formal criterion of kinetic equation is fulfilled only for $T\gtrsim 100$~K~\footnote{The possible origins of the scattering are: (i) short-range defects, (ii) exchange scattering with resident electrons. In the latter case for resident electron density $n_e \sim 10^{12}$~cm$^{-2}$ the exchange exciton-electron~\cite{tarasenko98} scattering yields $\tau \sim 10^{-13}$~s.}

For $4<T<30$~K, the measured exciton spin relaxation time is temperature independent (the sample temperature was carefully checked). In this temperature range, the PL spectra in Fig.~\ref{fig:fig1}b are also identical whereas shifts of the peaks energies and relative changes of intensities are clearly observed for larger temperatures. We believe that this behavior could be either due to (i) a regime where K$_BT$ is smaller than the collision broadening leading to a temperature independent spin relaxation time \cite{Glazov:2014a} or (ii) a localized character of the exciton below 30 K . 

Finally we emphasize that the Kerr rotation signal for delay times longer than $\sim$ 25 ps vanishes within our experimental accuracy whatever the temperature is. This means that we get no evidence of the transfer of spin polarization to the resident carriers though they are clearly present as shown by the detection of the trion line in Fig.~\ref{fig:fig1}. This behavior is probably due to the very robust spin/valley polarization of single particles, electrons and holes, in TMDC 2D layers \cite{Xiao:2012a}, because the polarization transfer from photogenerated carriers to resident ones \emph{n}-doped III-V or II-VI semiconductors usually requires a single particle spin-flip \cite{Glazov:2012a}.

In conclusion, we have demonstrated that the exciton valley dynamics in WSe$_2$ monolayer can be directly probed by time-resolved Kerr rotation dynamics performed in resonant excitation conditions. The temperature dependence of the exciton depolarization time is well described by the Coulomb exchange interaction induced exciton spin/valley dephasing assuming efficient scattering on short-range potential. 

This work was supported by the National Science Foundation of China Grant No. 11174338, Programme Investissements d'Avenir ANR-11-IDEX-0002-02, reference ANR-10-LABX-0037-NEXT and ERC Grant No. 306719.  X.M. acknowledges the support by the Chinese Academy of Sciences Visiting Professorship program for Senior International Scientists. Grant No. 2011T1J37. M.G. acknowledges the support by Russian Science Foundation (project 14-12-01067).

%
\end{document}